\documentclass[useAMS,usenatbib,usegraphicx]{mn2e}

\input{pjw_aas_macros.cls}
\newcommand{\asca}{{\it ASCA}}
\newcommand{\rosat}{{\it ROSAT}}
\newcommand{\absori}{{\it absori}}

\title[The X-ray spectrum of CH~Cyg]
{A new interpretation of the remarkable X-ray spectrum of the symbiotic star CH~Cyg}
\author[Peter J.\  Wheatley]{Peter J. Wheatley$^1$\thanks{E-mail: p.j.wheatley@warwick.ac.uk} and Timothy R.\ Kallman$^2$ \\
$^1$ Department of Physics, University of Warwick, Coventry CV4 7AL,
UK \\
$^2$ Laboratory for High-Energy Astrophysics, NASA Goddard Space Flight Center, Code 662, Greenbelt, MD 20771, USA}
\begin{document}

\date{}

\pagerange{\pageref{firstpage}--\pageref{lastpage}} \pubyear{2004}

\maketitle

\label{firstpage}

\begin{abstract}
We have reanalysed the \asca\ X-ray spectrum of the bright symbiotic star 
CH~Cyg, which exhibits apparently distinct hard and soft X-ray 
components. Our analysis demonstrates that the 
%distinct 
soft X-ray emission 
%component
can be interpreted as scattering of the hard X-ray component in a 
photo-ionised medium surrounding the white dwarf. 
This is in contrast to previous analyses in which the soft X-ray emission was 
fitted separately and assumed to arise independently of the hard X-ray 
component. 
%Several alternative emission mechanisms have been proposed to 
%explain the soft component, 
%including colliding winds, jet shocks and coronal emission from the giant 
%star. 
%A consequence of our analysis is that these mechanisms are 
%no longer required to explain the X-ray spectrum of CH~Cyg. 
%Finally, we 
We
note the striking similarity 
between the X-ray spectra of CH~Cyg and Seyfert 2 galaxies, which are also 
believed to exhibit scattering in a photo-ionised medium. 
\end{abstract}


\section{Introduction}
Symbiotic systems contain a red giant star but also exhibit high-ionisation
features in optical/ultraviolet spectra. In most cases it is believed that a 
white dwarf companion is accreting from the wind of the giant. 

Symbiotic stars tend also to be X-ray sources and \citet{Muerset97} used 
\rosat\ spectra to separate their X-ray emission into three distinct classes: 
$\alpha$) supersoft emission (below 0.4\,keV), $\beta$) emission with a 
characteristic temperature of a few $10^6$\,K (peaking at $\sim$0.8\,keV), 
and $\gamma$) harder X-ray emission. These classes were interpreted as arising 
from: $\alpha$) the photosphere of a hot white dwarf, $\beta$) the collision 
of the wind of the giant with a high-velocity wind from the hot companion, 
$\gamma$) a possible signature of a neutron star companion. 
One system, CH~Cyg, was classed as exhibiting both a $\beta$ component and a 
``residual $\gamma$ component''. 

\citet{Ezuka98} observed 
%the symbiotic system 
CH~Cyg with \asca , which was sensitive to harder X-rays than was \rosat\ 
(hard limits of 10\,keV and 2.5\,keV respectively). They found that the X-ray 
spectrum of CH~Cyg is made up of apparently distinct hard and soft X-ray 
components. The soft component peaks at 0.8\,keV and the hard component at 
5\,keV (see Fig.\,\ref{fig-spec}). \citet{Ezuka98} interpreted the hard 
component as heavily-absorbed optically-thin thermal emission of material 
being accreted by the white dwarf. In common with previous authors 
\citep{Leahy95,Muerset97} they fitted the soft emission as a distinct 
optically-thin emission component. 
This was interpreted as either coronal emission from the giant star or shock 
heating in the radio/X-ray jets of CH~Cyg. \citet{Leahy95} and 
\citet{Muerset97} both interpret the soft component as 
the result of shock heating in the collision of stellar winds from the 
giant and white dwarf. 

\citet{Wheatley03-4dra} reported a similar two-component 
spectrum in another symbiotic star, 4~Draconis. 
In this case the X-ray spectrum was seen to change between the two-component 
form and a single-component form. 
\citet{Wheatley03-4dra} interpreted both spectra as representing 
emission {\em only} from an accreting white dwarf, 
but with a varying amount of partially-ionised 
absorption. 
%Photoelectric absorption by neutral solar-abundance material 
%acts more strongly at softer energies. 
%However, ionisation 
Ionisation of the absorbing material removes absorption edges 
entirely from lighter elements (e.g. H, He), and shifts
absorption edges of heavier elements to higher energies (e.g. O),
allowing soft photons to leak through. 
%\citet{Wheatley03-4dra} showed that an apparantly two-component spectrum, 
%such as that of CH~Cyg, can be interpreted as..

In this paper we
%show that 
consider whether
similar models can account for the apparent 
two-component nature of the \asca\ spectrum of CH~Cyg. 
In doing so we draw attention to the striking similarity between the 
X-ray spectra of 
%4~Draconis, 
CH~Cyg and 
%the spectra of 
Seyfert 2 galaxies.

\section{Observations}
\label{sect-obs}
CH~Cyg was observed using the \asca\ X-ray observatory \citep{Tanaka94} 
on 1994 October 19. \asca\ carried four X-ray telescopes, of which two were 
equipped with CCD detectors (SIS0 and SIS1) and two with proportional counter 
detectors (GIS2 and GIS3). 
The SIS detectors had higher spectral resolution and better 
soft X-ray sensitivity, 
and in this paper we have limited our analysis to data from 
the SIS0 detector which also had the highest count rate.
We have used the standard screened 
data as extracted from the Leicester Database and Archive Service (LEDAS) at 
the University of Leicester. This yields an exposure time of 19.2\,ks in the 
SIS0 instrument.
% and 20.8\,ks in the GIS2 instrument. 

Data were reduced using {\sc ftools} software and standard techniques as 
described in the ASCA Data Reduction 
Guide\footnote{http://heasarc.gsfc.nasa.gov/docs/asca/abc/abc.html}. 
The SIS0 source spectrum was extracted from a circular region 
of radius 4 arcmin, and the background was estimated from most of 
the remainder of the CCD.
The mean background-subtracted count 
rate was $0.516\pm0.006\rm\,s^{-1}$, yielding $9890\pm120$ source
counts. A log of observations is presented in Table\,\ref{tab-log}.

\begin{table}
\begin{center}
\caption{Log of the \asca\ SIS0 observations presented in this paper. 
%of the Seyfert 2 galaxies presented in Fig.\,\ref{fig-sey2s}.
}
%\label{tab-sey2s}
\label{tab-log}
\begin{tabular}{ccccc}
Target & seq.\  no.& Date & Exposure & Count rate \\\hline
CH\,Cyg   & 42020000 & 1994-10-19 & 19.2\,ks & 0.516\,s$^{-1}$\\
NGC\,4507 & 71029000 & 1994-02-12 & 28.3\,ks & 0.126\,s$^{-1}$\\
NGC\,4388 & 73073000 & 1995-06-21 & 31.5\,ks & 0.032\,s$^{-1}$ \\
NGC\,7582 & 74026000 & 1996-11-21 & 39.7\,ks & 0.011\,s$^{-1}$ \\
\end{tabular}
\end{center}
\end{table}

\section{Spectral analysis}
\label{sect-spec}
\subsection{Two emission components?}
\label{sect-2compt}
\begin{figure}
\includegraphics[width=8.4cm]{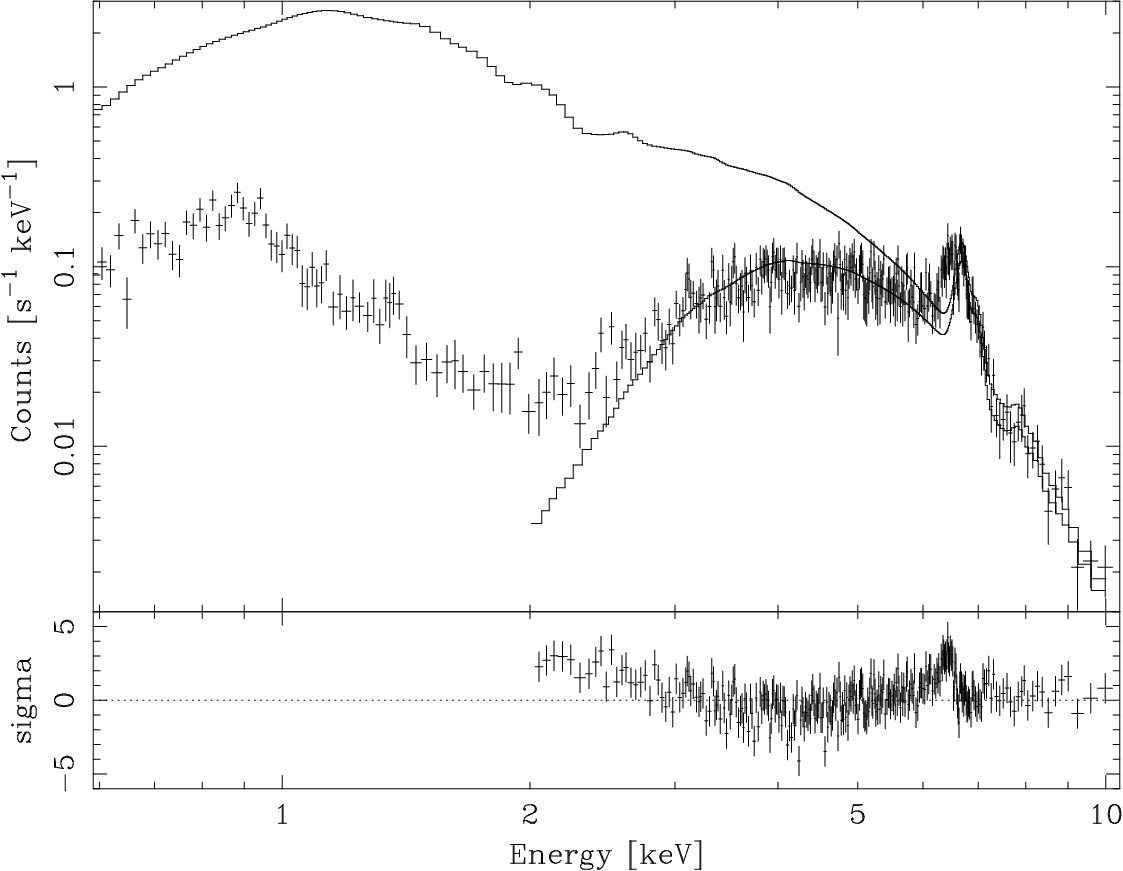}
\caption{\asca\ SIS0 spectrum of CH~Cyg. The solid line running through the 
data points represents a simple fit to the spectrum above 2\,keV (see text). 
The lower panel shows the residuals to the fit normalised by the error on 
each data point. The upper line in the top panel shows the unabsorbed model. }
\label{fig-spec}
\end{figure}

The SIS0 spectrum of CH~Cyg is presented in Fig.\,\ref{fig-spec}. 
The two X-ray components discussed by \citet{Ezuka98} are clearly visible 
either side of a minimum at 2\,keV. \citet{Ezuka98} fitted the spectrum with a 
complex 
multi-component model consisting of three separate plasma emission components 
(two for the soft emission and one for the hard emission) and three separate 
absorption components (one for the soft emission and two for the 
hard emission). 

The solid line in Fig.\,\ref{fig-spec} shows the results of a fit to the 
spectrum only above 2\,keV. We used the {\it mekal} optically-thin 
thermal plasma model \citep{Mewe85,Mewe86,Kaastra92,Liedahl95} 
absorbed by a neutral cosmic-abundance absorber. 
The best-fitting {\it mekal}
temperature was 7.7\,keV (characteristic of accreting white dwarfs) 
and the absorber column density was 
$1.6\times10^{23}\rm\,cm^{-2}$. As noted by \citet{Ezuka98}, this 
simple absorption model does not result in a good fit to the hard component
(we found a reduced $\chi^2$ of 1.9 with 282 degrees of freedom). This 
motivated \citet{Ezuka98} to add a second, complex absorption component to 
their hard X-ray model (they used a partial-covering model). They did not, 
however, consider the possibility that the soft component itself could arise 
as a consequence of 
%this complexity. 
%this complex absorption. 
photons leaking through or around this complex absorber. 
The upper line in Fig.\,\ref{fig-spec} shows the unabsorbed model spectrum. 
It can be seen that the implied 
soft X-ray flux from the hard component exceeds the detected flux by an 
order of magnitude. 
%Thus even
Even a small soft X-ray leak would easily account for 
the measured soft X-ray emission, avoiding the need for a second emission 
component. 

\subsection{A photo-ionised absorber?}
\label{sect-absori}
\citet{Wheatley03-4dra} successfully fitted the two-component X-ray 
spectrum of the 
symbiotic star 4~Draconis with a single emission component modified 
by a photo-ionised absorber. 
Ionisation of the absorbing medium removes absorption edges 
entirely from lighter elements (e.g.\  H, He) and shifts
absorption edges of heavier elements to higher energies (e.g.\  O).
%Ionisation of the absorbing medium 
%removes inner-shell electrons first from low-Z elements 
%that would otherwise dominate the absorption of soft X-rays. 
The result is that soft X-rays leak through the absorber, giving rise to what 
can appear to be a separate emission component. In this section we investigate
whether this same model might explain the two-component X-ray 
spectrum of CH~Cyg. 

In order to consider the possible effects of ionisation of the absorbing 
medium in CH~Cyg we employed the \absori\ model in {\sc xspec} 
\citep{Done92,Zdziarski95}. This is a simple single-zone photo-ionisation 
model which assumes a power-law input spectrum. 
For this paper we modified the \absori\ source code to use a 
bremsstrahlung spectrum as the ionising source, and we fixed the bremsstrahlung 
temperature to the {\it mekal} temperature of the hard component. 
%For the spectral fitting in 
%this section I fixed the bremstrahlung temperature to that of the {\it mekal} 
%emission component. 
The other 
parameters of the \absori\ model are the column density, $N_{\rm H}$, the 
temperature of the absorber, $T_{\rm abs}$, and the ionisation parameter
$\xi=L/nR^2$, where $L$ is the integrated source luminosity between 
5\,eV and 300\,keV, $n$ is the density of the material, and $R$ is the 
distance of the material from the illuminating source. 
In addition to the \absori\ model we included a neutral absorption component 
to account for interstellar absorption. 

\begin{figure}
\includegraphics[width=8.4cm]{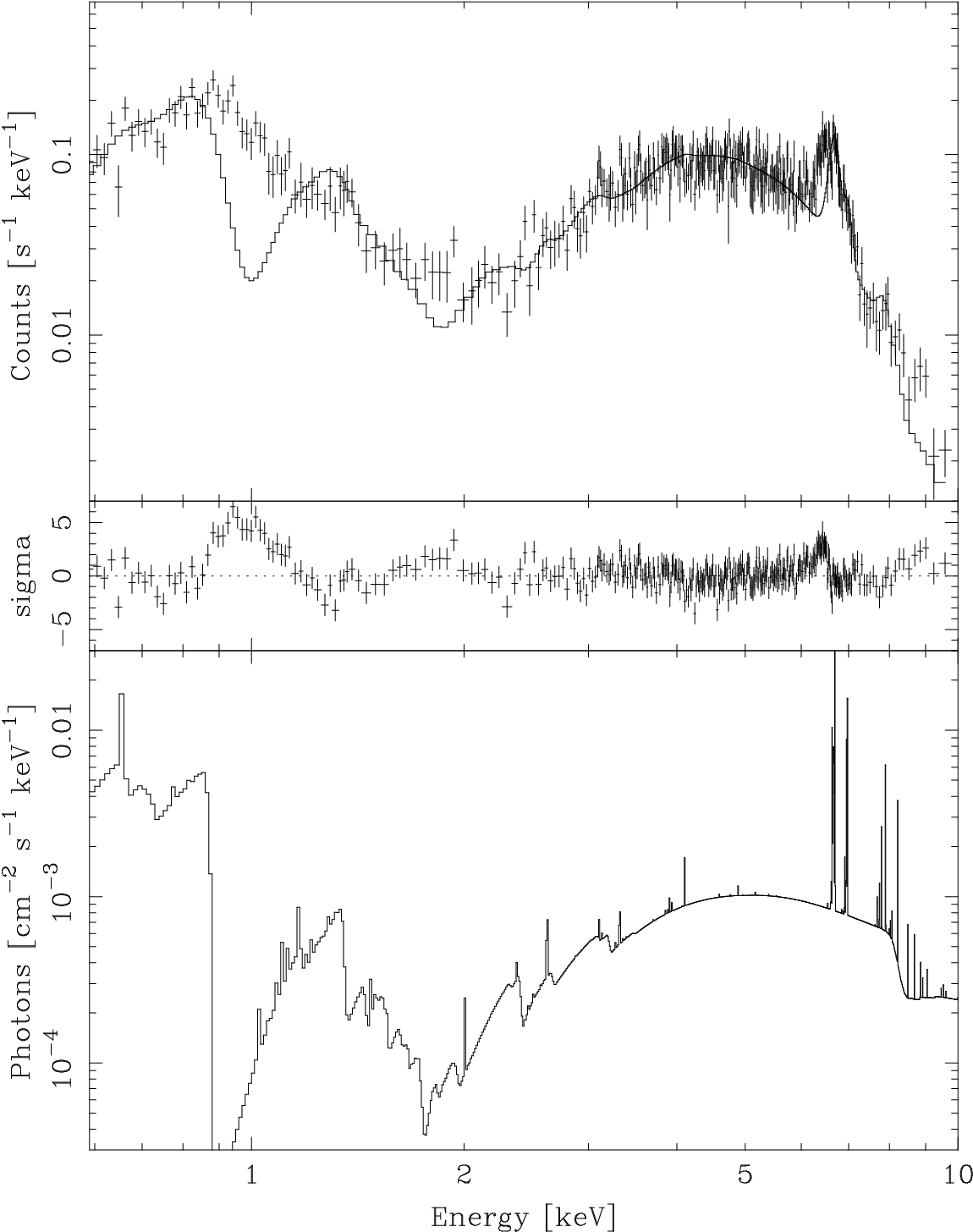}
\caption{\asca\ SIS0 spectrum of CH~Cyg fitted with the \absori\ ionised 
absorption model (top panel). The middle panel shows the residuals 
to the fit normalised by the error on each data point. The bottom panel shows 
the model spectrum. }
\label{fig-absori}
\end{figure}

Fitting with the \absori\ model we found we needed high values of $\xi$ 
($\sim10^2-10^3$) in order to allow the soft X-ray leak to extend as 
hard as 2\,keV. In contrast, the X-ray minimum between the two components 
occurred at around 1\,keV in 4~Draconis, and \citet{Wheatley03-4dra} found 
acceptable fits with $\xi\sim6$. 

The best fit with the \absori\ model is plotted in 
Fig.\,\ref{fig-absori} together with the model spectrum. 
It can be seen that the ionised absorber results in a better fit to the 
hard X-ray emission between 2--5\,keV than did the neutral absorber 
(Fig.\,\ref{fig-spec}).
It also allows a sufficiently strong soft-X-ray leak to account 
reasonably well for the soft X-ray emission 
(although the fit is not statistically acceptable with a reduced $\chi^2$ 
of 2.3 with 348 degrees of freedom).
The best-fitting column density for the 
ionised absorber is $N_{\rm H}=6.2\times10^{23}\rm\,cm^{-2}$. 
The ionisation parameter was found to be anti-correlated with the 
temperature of the absorber, and equally good fits were found 
for $\xi$ in the range 55--1000 (for $T_{\rm abs}=10^6-10^4$\,K). 
%and $\xi=660$. 
The best-fitting {\it mekal} 
temperature was 7.1\,keV
and the neutral absorber had a 
column density of $N_{\rm H}=2.2\times10^{21}\rm\,cm^{-2}$. 

Inspection of Fig.\,\ref{fig-absori} shows that the strongest
residuals 
occur at 
0.9--1.0\,keV, coincident with the strong OVIII K-shell edge in the \absori\ 
model. Residuals are also apparent around 6.4\,keV, 
the energy of the $K_\alpha$ line of neutral iron. 

Adding a narrow emission line fixed at 6.4\,keV entirely removed the
residuals at this energy and resulted in an improvement in reduced $\chi^2$ to 
1.94 (with 347 degrees of freedom). The line has an equivalent width of 
220\,eV, which is high for X-ray reflection from a cold slab
such as the white dwarf surface \citep{George91}. This indicates
either partial ionisation of the reflecting medium or fluorescence from an 
additional location, 
perhaps an accretion disc and/or the wind of the red giant.
%these numbers from paper_fit2.xcm

Removing the residuals in the 1\,keV region is more difficult. 
Adding a narrow line also in the 1\,keV region 
resulted in a considerable improvement in reduced 
$\chi^2$ to 1.22 (with 345 degrees of freedom) with a best-fitting line 
energy of 0.95\,keV. 
%these numbers from paper_fit3.xcm
This line might represent 
%thermal emission from relatively cool gas or 
%radiative recombination continuum emission from oxygen. 
recombination emission from oxygen, and indeed the 
fit statistic was further improved to 1.12 (345 d.o.f.) by replacing the 
narrow line with 
%an unabsorbed 
a
radiative recombination continuum (RRC) with 
edge energy fixed at the OVIII K-shell energy (0.8714\,keV). However, 
this fit relies on a high electron temperature (0.1\,keV) 
to make the RRC sufficiently broad to cancel out the OVIII absorption edge. 
The perfect cancelling of these two features leads us to believe that neither 
are really present in the spectrum of CH~Cyg. 

\subsubsection{Fluxes}
The absorbed 0.6--10\,keV flux of the best fit to the ionised absorber model 
was
$6.3\times10^{-11}\,\rm erg\,cm^{-2}\,s^{-1}$ and the implied unabsorbed 
flux of the {\it mekal} model was $2.0\times10^{-10}\,\rm erg\,cm^{-2}\,s^{-1}$
corresponding to a bolometric 
luminosity of $3\times10^{33}\,\rm erg\,s^{-1}$ at a 
distance of $270\pm65$\,pc \citep{Viotti97,Perryman97}.
This corresponds to an accretion rate onto a white dwarf of around 
$3\times10^{16}\,\rm g\,s^{-1}$ or
$5\times10^{-10}\,\rm M_{\sun}\,yr^{-1}$ (by considering the
gravitational potential energy of material falling from infinity). 
This is a fairly robust estimate since it depends mainly on the temperature 
and normalisation of the {\it mekal} component. It does not rely on the 
ionised absorber interpretation. 

\subsubsection{Implied geometry}
\label{sect-geom}
As stated above, the ionisation parameter is defined as $\xi=L/nR^2$. 
The column density is defined as $N_{\rm H}=n l$ where $l$ is the path length 
through the absorber. The ionisation parameter can be written, therefore,  
also as $\xi=Ll/N_{\rm H}R^2$. The path length through the absorber ($l$) 
cannot be much larger than the distance of the absorber from the ionising 
source ($R$), 
and so there is a maximum allowed distance of the absorber from the ionising 
source is given by $R_{\rm max}\sim L/N_{\rm H}\xi$. 
Taking the best fitting values for $L$ and $N_{\rm H}$ 
%and the lowest value for $\xi$ I find $R_{\rm max}\sim10^8\,\rm cm$, 
and the allowed range of $\xi$, we find 
$R_{\rm max}\sim5\times10^6-9\times10^7\,\rm cm$
which is much smaller than the size of a white dwarf 
(typically $\sim7\times10^8\,\rm cm$). Even allowing for 
%Therefore, even allowing for 
extra X-ray luminosity that might be hidden by the absorber, 
%an ionised absorber in CH~Cyg cannot be much larger than the white dwarf 
%itself. 
%would have to lie within a few white-dwarf radii of the white dwarf.
an ionised absorber in CH~Cyg 
would have to be located very close to the white dwarf. 
%This might represent the accretion flow itself, perhaps an 
%accretion disc or a magnetically-confined accretion curtain. 

\subsection{A photo-ionised scattering medium?}
\label{sect-xstar}
%\subsubsection{Motivation}
As well as soft X-rays leaking {\em through} the absorber, 
it is also possible for soft X-rays to leak {\em around} the absorber. This is 
possible if the absorber is not isotropically distributed around the white 
dwarf and instead absorbs preferentially along our line of sight. 
Soft photons emitted in our direction would then be 
%absorbed close to the white dwarf, 
absorbed, but soft photons emitted in other directions would 
escape further from the white dwarf where they could be scattered into 
our line of sight in a lower-density photo-ionised medium. 

This arrangement is essentially the same as is believed to exist in the 
nuclei of Seyfert 2 galaxies, where the anisotropic absorption is attributed 
to a highly inclined accretion disc or molecular torus. 
%CH~Cyg may be viewed at high inclination \citep[e.g.][]{Skopal96}, and 
%the presence of radio and X-ray jets \citep{Taylor86,Galloway04} 
%can be taken as evidence of an accretion disc.
The \asca\ spectra of Seyfert 2 galaxies are actually remarkably similar to 
the spectrum of CH~Cyg, and we discuss this further in 
Sect.\,\ref{sect-discuss}. 

In 
%However, in
order to 
consider the possible effects of photo-ionised scattering 
%show that the photo-ionised scattering model is at least plausible
in CH~Cyg
we have 
%employed model grids from 
calculated model grids using
the {\sc xstar} photo-ionisation code 
\citep{Kallman82, Kallman86, Turner96}. 
This code calculates the absorption and emission spectra of a spherical shell 
of material surrounding a source of ionising photons. 
We used the emission spectrum of a model grid covering an ionisation 
parameter ($\xi$) range of $10^{-4}-10^4$ 
(same definition as the \absori\ model above) and a column density range of 
$10^{19}-10^{23}\,\rm cm^{-2}$. The ionising spectrum was chosen to be an 
8\,keV bremsstrahlung in order to match approximately the emission of the 
white dwarf in CH~Cyg. This model grid (grid23scatt) is available 
from the {\sc xstar} 
webpage\footnote{http://heasarc.gsfc.nasa.gov/docs/software/xstar/xstar.html}.

\begin{figure}
\includegraphics[width=8.4cm]{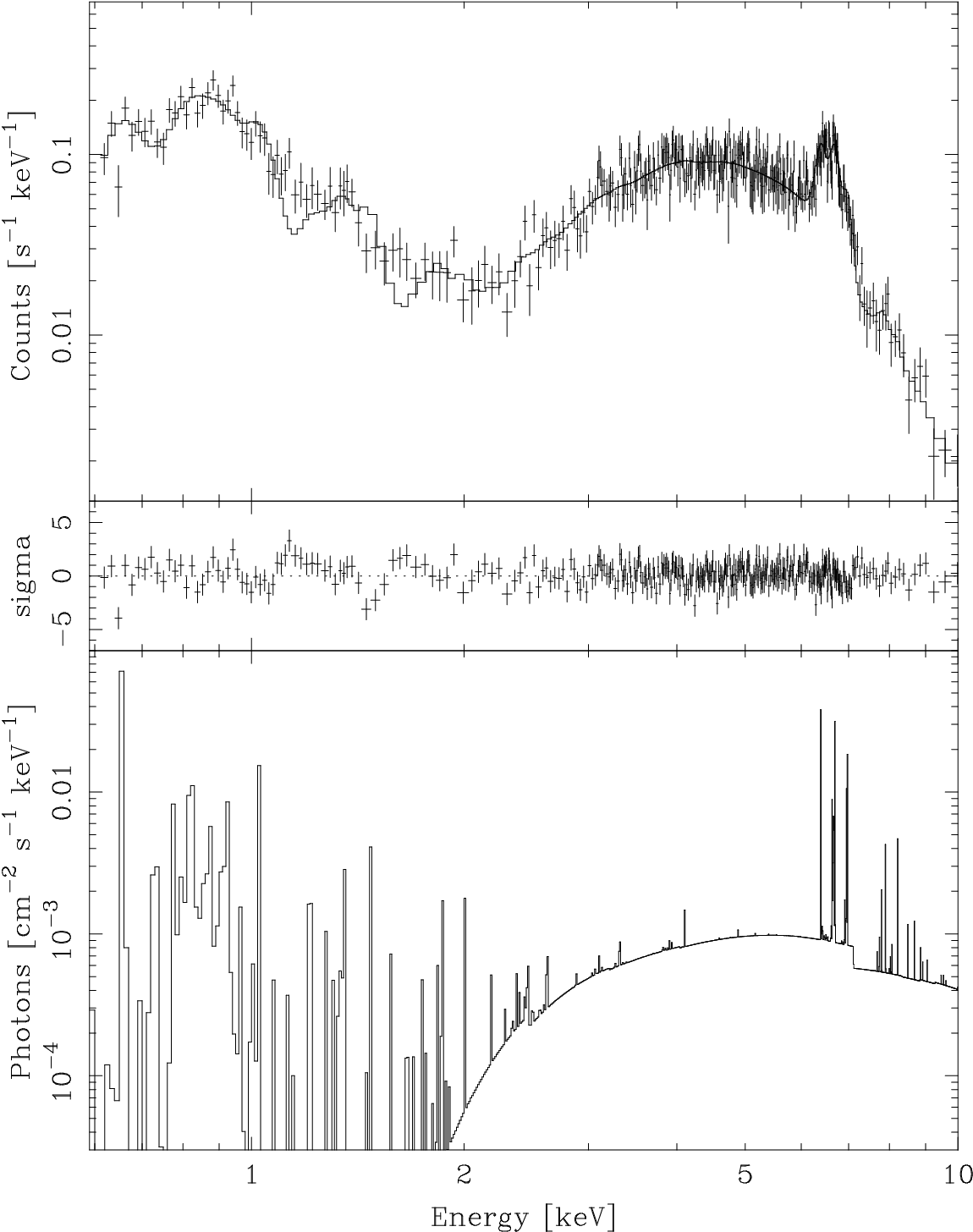}
\caption{\asca\ SIS0 spectrum of CH~Cyg fitted with {\sc xstar} photo-ionised 
scattering model (top panel). The middle panel shows the residuals to the fit 
normalised by the error on each data point. The bottom panel shows 
the model spectrum. }
\label{fig-xstar}
\end{figure}

We fixed the normalisation of the scattered spectrum to that appropriate to 
the unabsorbed flux of the fitted {\it mekal} emission 
component (a value of $4.1\times10^{-4}$). We replaced 
the ionised absorber model with a neutral absorber, but included a second 
partial-covering absorber in order to account for the complexity apparent in 
Fig.\,\ref{fig-spec}. We also included a narrow emission line 
fixed at 6.4\,keV and a third neutral absorber representing 
interstellar absorption.

Fitting with this model we found that the photo-ionised scattering region 
readily reproduced the strength and overall flux distribution of the 
soft component in CH~Cyg. We also found a statistically acceptable fit, with a 
reduced $\chi^2$ of 1.09 (338 d.o.f.). This best fit is plotted in 
Fig.\,\ref{fig-xstar}.

The strongest residuals in our best-fit scattering model occur between 
1--2\,keV. This corresponds to the energies of strong line emission in the 
{\it mekal} model, mainly due to iron L-shell, that were not included in our 
bremsstrahlung ionising spectrum. 
This missing input flux may account for these residuals. 
Alternatively, we have found that relaxing the constraint on the value of the 
normalisation of the scattering component allows a better fit to this region.
In effect this allows for extra ionising flux that is hidden by the main 
absorber, which is plausible because accretion by white dwarfs is known to 
result in emission at a wide range of temperatures \citep[e.g.][]{baskill05}. 
Finally, these residuals can also be reduced by allowing the relative 
elemental abundances to vary in the {\sc xstar} model. 
However, since individual emission features are not resolved, none of 
these possible explanations are unique, and we did not pursue them in detail. 
We were satisfied that the scattering model is capable of producing the 
strength and overall shape of the soft component in CH~Cyg. 

\subsubsection{Fitted parameters}
The fitted parameters of the scattering region were
$N_{\rm H}=(5.5\pm^{1.3}_{0.8})\times10^{20}\,\rm cm^{-2}$ and 
%$\log\xi=2.88\pm^{0.04}_{0.08}$. 
$\xi=760\pm^{70}_{130}$. 
The parameters for the {\it mekal} emission component were essentially 
unchanged, with $kT=8.3\pm0.4\,\rm keV$ and an unabsorbed 0.6-10\,keV flux of
$(2.29\pm0.15)\times10^{-10}\,\rm erg\,s^{-1}\,cm^{-2}$. The column densities 
of the main absorber are $(9.7\pm0.7)\times10^{22}\,\rm cm^{-2}$ and 
$(2.9\pm0.5)\times10^{23}\,\rm cm^{-2}$. The former was applied to all the 
flux from the hard component, the latter had a partial covering fraction 
of $66\pm4$ per cent. The column density of the absorber applied to the whole 
spectrum (including the scattered emission) was 
$(2.6\pm0.5)\times10^{21}\,\rm cm^{-2}$. The equivalent width of the 6.4\,keV 
line was $200\pm20$\,eV. 
All errors represent 68 per cent confidence 
intervals and account only for statistical errors.

\subsubsection{Implied geometry}
In Sect.\,\ref{sect-geom} we showed that our fits to the ionised absorber 
model 
placed tight constraints on the geometry of the absorber. Our scattering 
model relaxes these constraints because the absorption of the hard component 
is decoupled from the source of the soft component. Since the ionisation 
parameter of the absorber is consistent with zero, the absorber can be placed 
at any distance from the ionising source. However, the scattering model does 
require the absorption to be anisotropic around the white dwarf, and so the 
absorber is still likely to be closely associated with that object. 

The scattering region, in contrast, does have a well defined ionisation 
parameter in our model, and so we can apply the same 
argument used in Sect.\,\ref{sect-geom} to constrain its distance from the 
ionising source. In this case $R_{\rm max}\sim7\times10^9\,\rm cm$, which is 
around an order of magnitude larger than the white dwarf. 

%\section{Discussion}
\section{Discussion and conclusions}
\label{sect-discuss}
The spectral analysis of Sect.\,\ref{sect-spec} shows that the soft X-ray 
component in CH~Cyg can be interpreted as a reprocessing of the hard X-ray 
component. The X-ray spectrum is explained without the need for 
%There is no need for 
a second source of X-rays in the system, such as from colliding winds.

In Sect.\,\ref{sect-absori} we investigated the ionised absorber model 
used by \citet{Wheatley03-4dra}
to explain the two-component X-ray spectrum of 4~Draconis.
This model reproduced the strength of the observed soft component, but it 
predicted a strong OVIII absorption edge that is not detected. 
%This model did not provide an adequate explanation for the soft component, 
%because it required a highly-unlikely cancelling of OVIII absorption and 
%emission features. 
The required combination of high column density and high ionisation parameter
also implied that the absorber must be much smaller than the white dwarf.
%required that the absorber be much smaller than the 
%white dwarf, which we also regard as highly unlikely. 
Overall we do not regard this model as an adequate explanation of the soft 
X-ray component of CH~Cyg.

In Sect.\,\ref{sect-xstar} we considered an alternative explanation for 
the soft X-ray component in CH~Cyg: scattering of the hard component 
in a photo-ionised medium. In this interpretation the absorber is not 
isotropically distributed around the white dwarf and instead absorbs 
preferentially along our line of sight. 
Soft photons emitted in our direction are absorbed, but soft photons emitted 
in other directions escape further from the white dwarf where they can be 
scattered into our line of sight in a lower-density photo-ionised medium. 
This scattering model provided a much more satisfactory explanation for the 
soft component in CH Cyg. It readily reproduced the strength and shape of the 
soft component. It also implied a distance to the scattering region that is 
much larger than the white dwarf. 

\begin{figure}
\includegraphics[width=8.4cm]{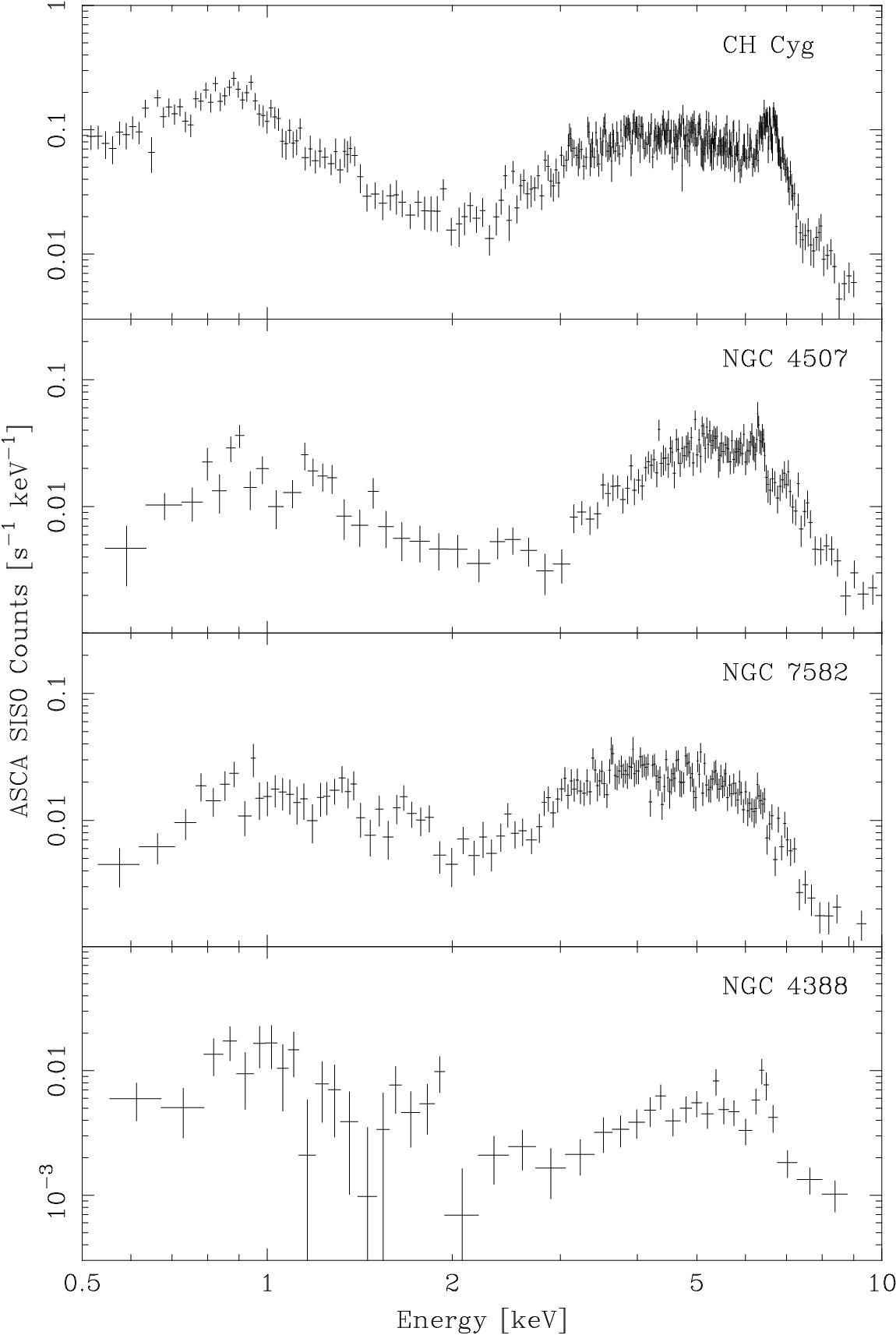}
\caption{\asca\ SIS0 spectrum of CH~Cyg compared with those of three Seyfert 2 galaxies. }
\label{fig-sey2s}
\end{figure}

As noted in Sect.\,\ref{sect-xstar}, the \asca\ spectrum of CH Cyg 
bears a striking resemblance to those of 
%(Compton-thin) 
Seyfert 2 active galaxies.
% do indeed bear a striking resembalance to that of CH~Cyg. 
Examples include NGC\,4388, NGC\,7582 and NGC\,4507 
\citep{Iwasawa97,Xue98,Comastri98}. 
In Fig.\,\ref{fig-sey2s} we plot the \asca\ SIS0 spectra of these systems, 
together with the spectrum of CH~Cyg. The Seyfert 2 spectra have been 
extracted using 
the same method as described for CH~Cyg in Sect.\,\ref{sect-obs}, and details 
of the observations are given in Table\,\ref{tab-log}.

In Seyfert 2 galaxies our line of sight to the central black hole is 
believed to pass through a highly inclined accretion disc, or molecular 
torus, that provides the strong X-ray absorption apparent in the \asca\ 
spectra (in some systems this region is Compton thick and blocks even 
hard X-rays). 
Soft X-ray components very much like that of CH~Cyg are also observed 
(Fig.\,\ref{fig-sey2s}), and this has generally been interpreted as 
scattering in photo-ionised cones above and below the accretion discs/tori. 
This interpretation was strengthened by the apparent detection of the same 
ionised material in absorption in Seyfert 1 galaxies. The geometry described 
here is illustrated by the schematic diagrams in 
Fig.\,7 of \citet{Awaki97} and Fig.\,6 of \citet{Kinkhabwala02}, and is 
essentially the same interpretation as we favour for CH~Cyg.
There has been some debate over whether the soft X-ray components of Seyfert 2 
galaxies can instead be dominated by collisionally ionised gas, 
perhaps from shock heating in outflows from the central region or from 
nuclear star-forming regions (similar interpretations to those 
proposed by previous authors for CH~Cyg). 
However, the scattering model has now been confirmed in the Seyfert galaxies 
Mrk\,3 and NGC\,1068 using high-resolution X-ray spectra from 
{\it Chandra} and {\it XMM-Newton} \citep{Sako00,Kinkhabwala02,Brinkman02}. 

%\begin{table}
%\begin{center}
%\caption{Log of the \asca\ SIS0 observations presented in this paper. 
%%of the Seyfert 2 galaxies presented in Fig.\,\ref{fig-sey2s}.
%}
%%\label{tab-sey2s}
%\label{tab-log}
%\begin{tabular}{ccccc}
%Target & seq.\  no.& Date & Exposure & Count rate \\\hline
%CH\,Cyg   & 42020000 & 1994-10-19 & 19.2\,ks & 0.516\,s$^{-1}$\\
%NGC\,4507 & 71029000 & 1994-02-12 & 28.3\,ks & 0.126\,s$^{-1}$\\
%NGC\,4388 & 73073000 & 1995-06-21 & 31.5\,ks & 0.032\,s$^{-1}$ \\
%NGC\,7582 & 74026000 & 1996-11-21 & 39.7\,ks & 0.011\,s$^{-1}$ \\
%\end{tabular}
%\end{center}
%\end{table}

The success of the scattering model in explaining the X-ray spectra of 
Seyfert 2 galaxies gives us increased confidence that this model is 
a likely explanation for the similar X-ray spectrum of CH~Cyg. 
In Seyfert 2 galaxies the source of absorption is believed to be an 
edge-on accretion disc or torus. A highly inclined accretion disc may 
also account for the observed absorption in CH~Cyg. The presence of a disc is 
supported by the detection of 
%both radio and X-ray 
jets 
\citep{Taylor86,Galloway04} and
there is 
further evidence that CH~Cyg may be viewed at high 
inclination \citep[e.g.][]{Skopal96}. 

The scattering interpretation of the X-ray spectrum of CH~Cyg is attractive 
because it requires only one source of X-rays: an accreting white dwarf. 
%The second X-ray component is explained 
%as either the soft leak of an absorber that we know must be present 
%(in order to explain the drop in X-rays between 5 and 2\,keV), 
%or as scattering in a lower-density region that is illuminated directly by 
%the white dwarf. 
In contrast, \citet{Ezuka98} invoke a second X-ray source, 
%with a similar brightness to the white dwarf, 
%{\em in addition} to a
and still require a complex 
treatment of absorption and emission (three components of each).
%\citet{Ezuka98} attribute their second X-ray source to a collision between 
%the wind of the giant and a fast wind from the accreting white dwarf. 
%A consequence of my new interpretation is that colliding winds are no longer 
%required in CH~Cyg, at least to explain the X-ray emission. 
Several mechanisms have been invoked in order to 
explain this second source of X-rays, including colliding winds, jet shocks 
and coronal emission from the giant star \citep{Leahy95,Muerset97,Ezuka98}. 
A consequence of our analysis is that these sources are 
no longer required to explain the X-ray spectrum of CH~Cyg. A definitive test 
of this model would be provided by 
%the search for RRCs in 
high-resolution X-ray grating spectra of CH~Cyg.

\section*{Acknowledgements}
We thank Koji Mukai, Simon Vaughan and Chris Done for useful discussions, and 
the Jeno Sokoloski for helpful comments on the original version of 
this paper. We also thank the referee, Jan-Uwe Ness, for helpful
comments. 
Data used in this paper have been obtained from the 
Leicester Database and Archive Service (LEDAS) at the University of
Leicester. 

\bibliographystyle{../../tex/mn/mn2e}
%\bibliographystyle{mn2e}
%\bibliography{/local/pjw/papers/tex/refs2}
\bibliography{../../tex/refs2}

\label{lastpage}

\end{document}